# Strain-engineering quantum anomalous Hall state in Janus MnBi$_2$Se$_x$Te$_{4-x}$ monolayers

Jiale Chen[1,2], Pengfei Li[2], Jun Hu[1*]

[1]Institute of High Pressure Physics, School of Physical Science and Technology, Ningbo University, Ningbo 315211, China.

[2]Key Laboratory of Materials Physics and Anhui Key Laboratory of Nanomaterials and Nanotechnology, Institute of Solid State Physics, Chinese Academy of Sciences, Hefei 230031, China

Exploring intrinsic magnetic topological insulators (TIs) for next-generation spintronic devices is still challenging in recent years. Here, we present a theoretical investigation on the electronic, magnetic and topological properties of monolayer (ML) Janus MnBi$_2$Se$_x$Te$_{4-x}$ (1≤x≤3), derived from two trivial magnetic semiconductors ML MnBi$_2$Se$_4$ and MnBi$_2$Te$_4$. Our band structure analysis reveals that two out of the six Janus structures exhibit band inversion induced by spin-orbit coupling. These structures are confirmed to have nonzero integer Chern numbers, indicating their topological nature. Moreover, the topological state is robust under moderate biaxial strains. Interestingly, applying compressive strain results in a high Chern number of 2 and enhances their magnetic stability at elevated temperatures. Our findings offer an effective strategy to engineer magnetic TI states within the ML MnBi$_2$Te$_4$ family.

[*] Email: hujun2@nbu.edu.cn

## I. Introduction

Topological insulators (TIs), extensively explored in the past decade, constitute an intriguing new class of condensed matter[1,2]. Their unique physical and chemical properties hold great promise for applications in various fields including electronics, spintronics, photoelectronics, catalysis, and more[1,2]. TIs are distinguished from conventional insulators by their quantized helical conducting edge states, which give rise to the quantum spin Hall effect (QSHE)[3,4]. The single-layer graphene was predicted as the first TI[5] ever since its discovery[6,7,8,9]. However, observing the QSHE in graphene is unfeasible due to its small TI gap which stems from the weak spin-orbit coupling (SOC)[10]. To enhance the SOC in graphene, decorating graphene by heavy adatoms was proposed[11,12]. Simultaneously, exploration of other TIs led to the experimental realization of TI states in HgTe quantum wells and $Bi_2Se_3$ family[13,14,15,16]. Nevertheless, intrinsic two-dimensional (2D) TIs with substantial TI gaps are still evasive[17].

When the time reversal symmetry of a TI is broken, the TI may turn into a Chern insulator, characterized by a nonzero integer Chern number[18,19,20]. Chern insulators, also known as magnetic TIs, exhibit topological properties that bring about the quantum anomalous Hall effect (QAHE), and their dissipationless boundary states provide a possible solution to the inevitable heating problem in large-scale integrated circuits. Initially, magnetic adatoms or dopants were considered effective sources of magnetization. For example, transition metal adatoms such as Fe, Co and W may result in the QAHE in graphene[21,22,23]; Mn-doped HgTe and (Cr or Fe)-doped $Bi_2Se_3$, $Bi_2Te_3$ and $Sb_2Te_3$ were predicted as magnetic TIs[18,24], with experimental observation of the QAHE in Cr-doped $(Bi,Sb)_2Te_3$ later[25]. However, such approaches often lead to an inhomogeneous distribution of magnetic moments, making it difficult to control the electronic and magnetic properties in these magnetic TIs. Accordingly, intrinsic Chern insulators are desired, because the magnetic moments form periodic lattices, providing opportunities to investigate the intertwined band topology and magnetic order. Nonetheless, intrinsic magnetic TIs remained elusive until the topological properties of multilayer van der Waals $MnBi_2Te_4$ were unveiled[26,27]. Notably, the QAHE only manifests in odd-layer (except single-layer) $MnBi_2Te_4$, where the intra-layer and interlayer exchange coupling is respectively ferromagnetic and antiferromagnetic, while single-layer or monolayer (ML) $MnBi_2Te_4$ behaves as a normal insulator[28,29]. Accordingly, it is particularly interesting if the QAHE can be realized in a pure 2D ML of $MnBi_2Te_4$ or related materials due to their promising applications in next-generation spintronic devices[30]. In fact, numerous 2D materials exhibiting the intriguing QAHE have been theoretically predicted in recent years. For instance, ferromagnetic or antiferromagnetic monolayer oxides, such as $Nb_2O_3$ and MoO, have been identified as promising candidates to realize the QAHE at temperatures above room temperature[31,32]. In addition, 2D transition-metal chlorides, including $FeCl_2$ and RuClBr, demonstrate both the QAHE and valley Hall effect[33,34]. These advances further inspire the exploration of 2D forms that may exhibit the QAHE within the $MnBi_2Te_4$ family.

The Janus structure presents an interesting avenue for manipulating the electronic properties of 2D van der Waals materials[35]. In a Janus material, two different elements from the same group terminate the opposing sides, resulting in uniform electric dipoles between them due to their disparate electronegativities. This alignment of electric dipoles establishes an inner electric field along the surface normal of the Janus material, which tunes the electronic and magnetic properties. A prominent example of Janus materials is ML MoSSe, where a layer of S atoms is substituted with Se atoms[36,37]. Consequently, ML MoSSe exhibits significantly distinct physical and chemical characteristics from ML $MoS_2$ and $MoSe_2$ due to the inner electric field and the breaking of inversion symmetry. Similarly, introducing the Janus structure to ML $MnBi_2Te_4$ is expected to modify its electronic structure. In fact, piezoelectric ferromagnetism has been predicted theoretically in ML Janus $MnSbBiTe_4$[38]. On the other hand, theoretical calculations suggest that the members of the $MnBi_2Te_4$ family, i.e. $MnBi_2Se_4$, $MnSb_2Te_4$ and $MnSb_2Se_4$, could potentially manifest as magnetic TIs in their multilayer forms under specific conditions[39,40,41,42], and the topological properties of $MnBi_2Se_4$ have been demonstrated in experiment[43]. Therefore, exploring Janus derivatives of ML $MnBi_2Te_4$, such as $MnBi_2Se_xTe_{4-x}$, holds promise for uncovering appealing electronic and magnetic properties.

In this work, we first explored the electronic and magnetic properties of ML $MnBi_2Se_4$ and $MnBi_2Te_4$ using first-principles calculations, confirming that they are trivial magnetic semiconductors without any band inversion around their band gaps. By constructing Janus structures based on these maternal materials, we observed band inversion in two out of eight candidates. Calculations of their Hall conductance confirmed that both materials are Chern insulators. Furthermore, their topological nature remains robust under moderate biaxial strains, and a high Chern number of 2 can be achieved under compressive strains. Additionally, the compressive strain enhances their magnetic stability at elevated temperatures. Our work provides an effective approach to engineer Chern insulators within the family of ML MnBi$_2$Te$_4$.

## II. Computational details

First-principles calculations were performed within the density functional theory (DFT) framework using the Vienna *ab* initio Simulation Package (VASP)[44,45]. The interaction between valence electrons and ionic cores was described by the projector augmented wave (PAW) method[46,47], and electronic wavefunctions were expanded using a plane-wave basis set. A kinetic energy cutoff of 500 eV was applied for the plane-wave expansions. The exchange-correlation potential was treated at the generalized gradient approximation (GGA) level with the Perdew–Burke–Ernzerhof (PBE) functional[48]. The SOC effect was incorporated through a second variational procedure on a fully self-consistent basis. Convergence criteria were set at 0.01 eV/Å for forces and 10$^{-5}$ eV for total energy. Additionally, a Hubbard U correction was introduced to the Mn-3d orbitals to account for strong correlation effects[49], and a value of 4 eV for U was adopted, as it accurately captures the key physical properties of ML MnBi$_2$Se$_4$ and MnBi$_2$Te$_4$, such the lattice constants and magnetic moments. A Γ-centered Monkhorst-Pack k-point mesh of 24×24 was utilized in our calculations. To analyze the topological properties of the calculated materials, we employed the tight-binding model developed by Wu et al.[50], which is based on hopping parameters derived from the Wannier90 code[51].

## III. Results and discussion

In ML MnBi$_2$Te$_4$, septuple layers of atoms are stacked in the sequence of Te-Bi-Te-Mn-Te-Bi-Te, as illustrated in Fig. 1. The Te atoms within the inner and outer layers are denoted as Te' and Te", respectively. Each Te" atom coordinates with three Bi atoms, while all other atoms are surrounded by six neighbors that constitute distorted octahedra. ML MnBi$_2$Se$_4$ shares the same crystal structure, with Se atoms replacing Te atoms. We first optimized the lattice constants (*a*) of ML MnBi$_2$Se$_4$ and MnBi$_2$Te$_4$, yielding 4.12 and 4.37 Å, respectively. These values are slightly larger than the experimental measurements[43,52] but are consistent with other theoretical calculations[40,41,42]. Both ML MnBi$_2$Se$_4$ and MnBi$_2$Te$_4$ exhibit ferromagnetic ground state, with a total spin moment of 5 $\mu_B$ per unit cell, and each Mn atom contributing more than 4.8 $\mu_B$.

The band structures of ML MnBi$_2$Se$_4$ and MnBi$_2$Te$_4$ are plotted in Fig. 2. In the absence of SOC, both materials exhibit similar band structures, featuring indirect band gaps of 0.82 and 0.90 eV, respectively, as shown in Fig. 2(a) and 2(e). It can further be seen that the band gap predominantly arises from the majority-spin channel, with the conduction band minimum (CBM) located at the Γ point and the valence band maximum (VBM) along the path from Γ to K. To elucidate the characteristics of bands near the Fermi level ($E_F$), we analyzed contributions of atomic orbitals to the corresponding wavefunctions. As depicted in Fig. 2(b) and 2(f), the top valence bands are mostly contributed from the Se/Te-$p_{x/y}$ orbital, whereas the bottom conduction bands are primarily derived from the Bi-$p_z$ orbital.

Given that Se, Te and Bi are heavy elements, substantial SOC effects are anticipated to influence the band structures, especially the bands near the band gap. Therefore, we computed the band structures incorporating SOC, as presented in Fig. 2(c) and 2(g). Evidently, the inclusion of SOC leads to significant band splitting for both ML MnBi$_2$Se$_4$ and MnBi$_2$Te$_4$, resulting in a remarkable reduction in band gaps to 0.43 eV and 0.25 eV, respectively. Obviously, the reduction in ML MnBi$_2$Te$_4$ is more pronounced, attributable not only to the larger SOC constant of Te compared to Se but also to the transition of its band gap from indirect to direct. The band contributing to this transition is indicated by arrows in Fig. 2(f) and 2(h). Furthermore, the states near the VBM are dominantly contributed from the Se/Te-$p_{x/y}$ orbital, while those near the CBM exhibit a hybridization of the Bi-$p_z$ and Se/Te-$p_{x/y}$ orbitals, with the former being more predominant. It's worth noting that band inversion, typically induced by the SOC effect, serves as a hallmark for demonstrating the presence of topological phases[23,24]. However, band inversion is not observed in ML MnBi$_2$Se$_4$ and MnBi$_2$Te$_4$, as shown in Fig. 2(d) and 2(h), agreeing with previous reports[28,29].

We then investigated the structural and electronic properties of ML Janus MnBi$_2$Se$_x$Te$_{4-x}$. Starting from ML MnBi$_2$Te$_4$ (x = 0), we replaced one entire Te layer with Se atoms each time. As shown in Fig. 1, the Se layer can substitute either the Te' layer or the Te'' layer on both sides of the Mn layer, respectively marked as Se' and Se''. Consequently, two configurations arise for x = 1, MnBi$_2$Se'Te$_3$ and MnBi$_2$Se''Te$_3$, owing to the inversion symmetry of ML MnBi$_2$Te$_4$. Similarly, for x = 3, we also have two configurations: MnBi$_2$Se$_3$Te' and MnBi$_2$Se$_3$Te''. For x = 2, four distinct configurations can be constructed with two Se layers replacing two different Te layers: both Te'' layers (designated as MnBi$_2$Se$_2$Te$_2$-I), one Te' layer and one Te'' layer on opposite sides of the Mn layer (MnBi$_2$Se$_2$Te$_2$-II), one Te' layer and one Te'' layer on the same side of the Mn layer (MnBi$_2$Se$_2$Te$_2$-III), both Te' layers (MnBi$_2$Se$_2$Te$_2$-IV). Note that MnBi$_2$Se$_2$Te$_2$-I and MnBi$_2$Se$_2$Te$_2$-IV are not Janus structure due to the symmetric arrangement of the Se and Te layers with respect to the Mn layer. Consequently, six Janus structures are identified among the eight ML MnBi$_2$Se$_x$Te$_{4-x}$ (1≤x≤3). All the structures were fully optimized, from which the lattice constants gradually vary from that of ML MnBi$_2$Te$_4$ (4.37 Å) to that of ML MnBi$_2$Se$_4$ (4.12 Å).

We found that the band structures of ML Janus MnBi$_2$Se$_x$Te$_{4-x}$ are indeed significantly different from those of both ML MnBi$_2$Se$_4$ and MnBi$_2$Te$_4$. Especially, their band gaps change drastically with respect to their maternal materials, and are no longer between those of ML MnBi$_2$Se$_4$ and MnBi$_2$Te$_4$. As seen in the inset in Fig. 3, the band gaps of six ML MnBi$_2$Se$_x$Te$_{4-x}$ are smaller than those of both ML MnBi$_2$Se$_4$ and MnBi$_2$Te$_4$, while the band gaps of other two ML MnBi$_2$Se$_x$Te$_{4-x}$ are smaller than that of ML MnBi$_2$Se$_4$ but larger than that of ML MnBi$_2$Te$_4$. In addition, the band gaps of three cases with Se layers on Mn sides (i.e. MnBi$_2$Se$_3$Te'', MnBi$_2$Se$_2$Te$_2$-IV and MnBi$_2$Se$_4$) are clearly larger than those of the others. However, only two cases, MnBi$_2$Se$_2$Te$_2$-III and MnBi$_2$Se$_3$Te', exhibit band inversion which will be discussed later, indicating possible topological insulator state.

To investigate the stability of the ML MnBi$_2$Se$_2$Te$_2$-III and MnBi$_2$Se$_3$Te', we first calculated the phonon spectra of MnBi$_2$Se$_2$Te$_2$-III and MnBi$_2$Se$_3$Te', as presented in Fig. 3(b) and 3(c). It is evident that the atomic structure of MnBi$_2$Se$_2$Te$_2$-III is dynamically stable, because all phonon frequencies are positive. For MnBi$_2$Se$_3$Te', however, a small portion of the acoustic branch shows negative frequencies near the Γ point. Nevertheless, the magnitudes of these negative frequencies are quite small, suggesting that the atomic structure should remain stable under ambient environment or when placed on a substrate. We further calculated the elastic constants $C_{ij}$ of the ML MnBi$_2$Se$_2$Te$_2$-III and MnBi$_2$Se$_3$Te' to examine their mechanical stability. Considering the symmetry of the 2D hexagonal lattice, three independent elastic constants ($C_{11}$, $C_{12}$, $C_{66}$) were obtained, namely $C_{11}$ = 82.13 N/m, $C_{12}$ = 23.65 N/m, and $C_{66}$ = 29.24 N/m for MnBi$_2$Se$_2$Te$_2$-III, while $C_{11}$ = 86.27 N/m, $C_{12}$ = 25.53 N/m, and $C_{66}$ = 30.3698 N/m for MnBi$_2$Se$_3$Te'. Clearly, these values satisfy the Born−Huang criteria ($C_{11}C_{22} - C_{12}^2 >$ 0 and $C_{66} > 0$)[53], indicating good mechanical stability. On the other hand, Janus structures of various 2D materials have been extensively studied in experiment[35,36,37], demonstrating the feasibility of synthesizing ML MnBi$_2$Se$_2$Te$_2$-III and MnBi$_2$Se$_3$Te' as proposed in this work.

The spin-polarized band structures, depicted in Fig. 4(a) and 4(e), reveal that the ML

MnBi$_2$Se$_2$Te$_2$-III and MnBi$_2$Se$_3$Te' exhibit band gaps larger than 500 meV, with both the VBM and CBM originating from the spin-majority states. Additionally, the VBM and CBM are almost exclusively contributed by the Te-p$_{x/y}$ and Bi-p$_z$ orbitals, respectively, as shown in Fig. 4(b) and 4(f). After the SOC effect is included, the band gaps undergo a notable reduction, as seen in Fig. 4(c) and 4(g), down to merely 40.3 meV and 53.7 meV for MnBi$_2$Se$_2$Te$_2$-III and MnBi$_2$Se$_3$Te', respectively. Most intriguingly, a band inversion emerges between the Te-p$_{x/y}$ and Bi-p$_z$ states near the Γ point in both materials, as indicated by the orbital-resolved band structures in Fig. 4(d) and 4(h). Furthermore, both the VBM and CBM of ML MnBi$_2$Se$_2$Te$_2$-III are located at the Γ point, while those of ML MnBi$_2$Se$_3$Te' slightly shift away from the Γ point due to the intersection between the topmost valence band and lowest conduction band.

Since biaxial strain (ε) is an effective way to engineer the electronic structures of 2D materials[54], we applied biaxial strain from -5% to 5% to ML MnBi$_2$Se$_2$Te$_2$-III and MnBi$_2$Se$_3$Te', as well as ML MnBi$_2$Se$_4$ and MnBi$_2$Te$_4$ for comparison, to examine the effects of biaxial strain on their electronic structures. As seen in Fig. 3, the band gap of ML MnBi$_2$Se$_4$ linearly depends on the biaxial strain, monotonically decreasing as the strain varies from -5% to 5%. For ML MnBi$_2$Te$_4$, the band gap decreases for both the compressive and tensile strains. The band gap ML MnBi$_2$Se$_2$Te$_2$-III increases under compressive strain to 81 meV at a strain of -5%, whereas under tensile strain, its band gap decreases to 4.3 meV initially at a strain of 1% and then increases to 114 meV at s strain of 5%. The band gap of ML MnBi$_2$Se$_3$Te' is nonmonotonic under both compressive and tensile strains, varying between 9.4 meV and 58 meV.

Although the band structures of ML MnBi$_2$Se$_4$ and MnBi$_2$Te$_4$ are significantly modified by strain, no band inversion occurs, so they remain normal semiconductors. For ML MnBi$_2$Se$_2$Te$_2$-III and MnBi$_2$Se$_3$Te', the band inversion conserves under both compressive strain and moderate tensile strain (up to 1% and 3%, respectively) as depicted in Fig. 5. Moreover, the evolution of the bands under strain for these two cases is similar. Nevertheless, the valence band and conduction band of ML MnBi$_2$Se$_3$Te' intersect and reopen under compressive strain between -2% and -3%, causing the magnitude of the band gap to display a nonmonotonic dependence on the compressive strain as shown in Fig. 3.

It is known that a characteristic of a 2D magnetic TI is the quantized anomalous Hall conductance, i.e. the QAHE, given by σ$_{xy}$ = $C$ (e$^2$/h) with nonzero integer $C$. In this expression, $C$ represents the Chern number, which is determined by integrating the

Berry curvature, $\Omega(\mathbf{k})$, over the 2D Brillouin zone (BZ) as[23]

$$C = \frac{1}{2\pi}\int_{BZ}\Omega(\mathbf{k})d\mathbf{k}, \quad (1)$$

with $\Omega(\mathbf{k})$ being expressed as

$$\Omega(\mathbf{k}) = -2\,\text{Im}\sum_{n\in\{o\}}\sum_{m\in\{u\}}\frac{\langle\psi_{n\mathbf{k}}|v_x|\psi_{m\mathbf{k}}\rangle\langle\psi_{m\mathbf{k}}|v_y|\psi_{n\mathbf{k}}\rangle}{(\varepsilon_{m\mathbf{k}}-\varepsilon_{n\mathbf{k}})^2}, \quad (2)$$

where {o} and {u} stand for the sets of occupied and unoccupied states, respectively; $\psi_{n\mathbf{k}}$ and $\varepsilon_{n\mathbf{k}}$ are the spinor Bloch wave function and eigenvalue of the $n$th state at $\mathbf{k}$ point; and $v_{x(y)}$ is the velocity operator.

We first calculated the $E_F$-dependent $\sigma_{xy}$ of ML MnBi$_2$Se$_4$ and MnBi$_2$Te$_4$, as plotted in Fig. 6(a). The $\sigma_{xy}$ curves clearly display zero values near $E_F$, which further confirms their topologically trivial semiconducting nature. Subsequently, we calculated the $E_F$-dependent $\sigma_{xy}$ of ML MnBi$_2$Se$_2$Te$_2$-III and MnBi$_2$Se$_3$Te' under strain ranging from -5% to 5%. The results indicate that the cases without band inversion are indeed topologically trivial semiconductors, as evidenced by the zero $\sigma_{xy}$ within their band gaps. For the cases with band inversion, the $E_F$-dependent $\sigma_{xy}$ curves are presented in Fig. 6(b) and 6(c), covering strains of -5% to 1% for ML MnBi$_2$Se$_2$Te$_2$-III and -5% to 3% for ML MnBi$_2$Se$_3$Te'. Notably, most of these cases exhibit a plateau around $E_F$, associated with an integer Chern number of either 2 or -1. The width of these plateaus corresponds to their band gaps. Apparently, the nonzero integer Chern numbers definitely demonstrate the topological feature of these materials. Moreover, some cases, such as -5% to -4% for ML MnBi$_2$Se$_2$Te$_2$-III and 0% to 2% for ML MnBi$_2$Se$_3$Te', maintain relatively large band gaps, which is beneficial for realizing the QAHE at elevated temperatures.

It is interesting to reveal the underlying mechanism behind the variation in Chern numbers. As defined in Eq. (1), the Chern number corresponds to the integral of the Berry curvature over the Brillouin zone. When strain is applied to the material, the band structure is modified, which in turn affects the Berry curvature, as expressed in Eq. (2). Figure 7 plots the $k$-dependent Berry curvature [$\Omega(k)$] of MnBi$_2$Se$_2$Te$_2$-III under compressive strains of -5% and -2%, as well as strain-free case (0%). It is evident that significant values of $\Omega$ are concentrated near the $\Gamma$ point. For strains of -5% and 0%, the magnitudes of $\Omega$ are large but exhibit opposite signs, resulting in opposite integer Chern numbers. In contrast, for a strain of -2%, $\Omega$ remains positive but with a much smaller amplitude compared to the strain-free case. Consequently, the strain of -2% yields a positive non-integer Chern number. Similarly, the alteration of Chern numbers in ML MnBi$_2$Se$_3$Te' under strains are also closely associated with the change in Berry curvature.

It is important to note that the sign of a Chern number indicates the chirality, whereas its magnitude reflects the number of topologically protected chiral edge channels.[55] Most magnetic TIs exhibit an intrinsic Chern number of 1, implying that there is only one conducting channel at their edges. This limitation affects the efficiency of energy conservation when magnetic TIs are applied in spintronic devices[25,56]. Therefore, magnetic TIs with higher Chern numbers are desired to achieve more dissipationless chiral edge states. It has been reported that stacking nine or ten monolayers of MnBi$_2$Te$_4$ results in a Chern number of 2 under a relatively large magnetic field[57]. Additionally, by engineering repeat magnetic-TI/normal-insulator interfaces, the Chern number can be tuned from 1 to 5 depending on the thickness of the multilayer structures[56]. Intriguingly, a Chern number of 2 can be obtained in ML MnBi$_2$Se$_2$Te$_2$-III and MnBi$_2$Se$_3$Te' under a compressive strain of -4% to -5%. This discovery not only holds promise for practical applications but also provides a novel approach to tune the Chern number.

It is well known that the hybrid HSE functional is widely used to obtain band gaps of semiconductors comparable to experimental measurements. After conducting a series of calculations within the framework of the hybrid HSE06 functional using different parameters, we found the band gaps of ML MnBi$_2$Se$_2$Te$_2$-III and MnBi$_2$Se$_3$Te' are highly sensitive to the mixing parameter (α) of the Hartree-Fock exchange. Taking ML MnBi$_2$Se$_2$Te$_2$-III as an example, its band gap was calculated to be 391 and 118 meV, respectively for α = 0.25 and 0.1. In both calculations, the band inversion that characterizes the TI state disappears due to excessively large band gaps. However, a calculation with α = 0.05 yields a band gap of 38 meV, where the band inversion emerges. Considering that the Hubbard U correction also enhances exchange energy, we carried out additional HSE06 calculations with U = 0 to eliminate the double-counting of exchange energy. Under these conditions, calculations with α = 0.25 and 0.1 yield band gaps of 291 and 22 meV, respectively. As expected, the band inversion disappears in the former case but reappear in the latter. We further increased α to 0.15 and found that the band inversion still persists, confirming that the TI state is indeed influenced by the magnitude of the band gap. Given that TI phases have been experimentally observed in multilayer MnBi$_2$Te$_4$, we propose that the HSE band gap should not be used as a definitive criterion for the existence of TI states. A thorough evaluations of the relationship between HSE band gaps and TI states would require more intricate studies, which are beyond the scope of the present work.

Another critical factor for realizing the QAHE at elevated temperatures is the magnetic anisotropy of magnetic TIs. Accordingly, we calculated the magnetic anisotropy energies (MAEs) of ML $MnBi_2Se_4$, $MnBi_2Te_4$, $MnBi_2Se_2Te_2$-III and $MnBi_2Se_3Te'$ using the torque method[58,59]. As shown in Fig. 7, all four materials exhibit positive MAEs without strain, indicating intrinsic perpendicular magnetic anisotropy, agreeing with experimental observations[28,29,43]. Notably, ML $MnBi_2Se_2Te_2$-III possesses a much larger MAE (0.5 meV per unit cell) than the other cases (< 0.1 meV per unit cell), indicating a excellent intrinsic magnetic stability. Interestingly, the MAEs of ML $MnBi_2Se_2Te_2$-III and $MnBi_2Se_3Te'$ remarkably increase under compressive strain, to 0.95 and 0.83 meV, respectively, when the compressive strain reaches -5%. Therefore, the Janus structures of ML $MnBi_2Se_2Te_2$-III and $MnBi_2Se_3Te'$ not only lead to a transition from trivial magnetic insulators to Chern insulators, but also greatly enhances the stability of magnetization at elevated temperatures. This enhancement is mainly attributed to the states near the Γ point in the Brillouin zone, as illustrated in the inset in Fig. 7.

**IV. Conclusions**

In summary, we systematically investigated the electronic, magnetic and topological properties of ML $MnBi_2Se_4$, $MnBi_2Te_4$ and Janus $MnBi_2Te_xSe_{4-x}$ using first-principles calculations. We found that while $MnBi_2Se_4$ and $MnBi_2Te_4$ are trivial semiconductors, engineering the Janus structure $MnBi_2Te_xSe_{4-x}$ based on them results in a band inversion phenomenon in ML $MnBi_2Se_2Te_2$-III and $MnBi_2Se_3Te'$. Analysis of their band structures reveals that this band inversion, induced by SOC, occurs between the Bi-$p_z$ and Te-$p_{x/y}$ orbitals, which is a preliminary characteristic of TIs. In addition, this band inversion is robust under moderate biaxial strains. Further calculations of the Hall conductance of ML $MnBi_2Se_2Te_2$-III and $MnBi_2Se_3Te'$ reveal that both exhibit nonzero integer Chern number, confirming their nature as magnetic TIs. This topological feature is preserved under moderate biaxial strains. Notably, a high Chern number of 2 was achieved in ML $MnBi_2Se_2Te_2$-III and $MnBi_2Se_3Te'$ under a compressive strain of -4% to -5%. Additionally, the magnetocrystalline anisotropy of both monolayers are remarkably enhanced under compressive strain, improving its magnetic stability at elevated temperatures.

## Acknowledgements

This work is supported by the Program for Science and Technology Innovation Team in Zhejiang (Grant No. 2021R01004), the start-up funding of Ningbo University and Yongjiang Recruitment Project (432200942).

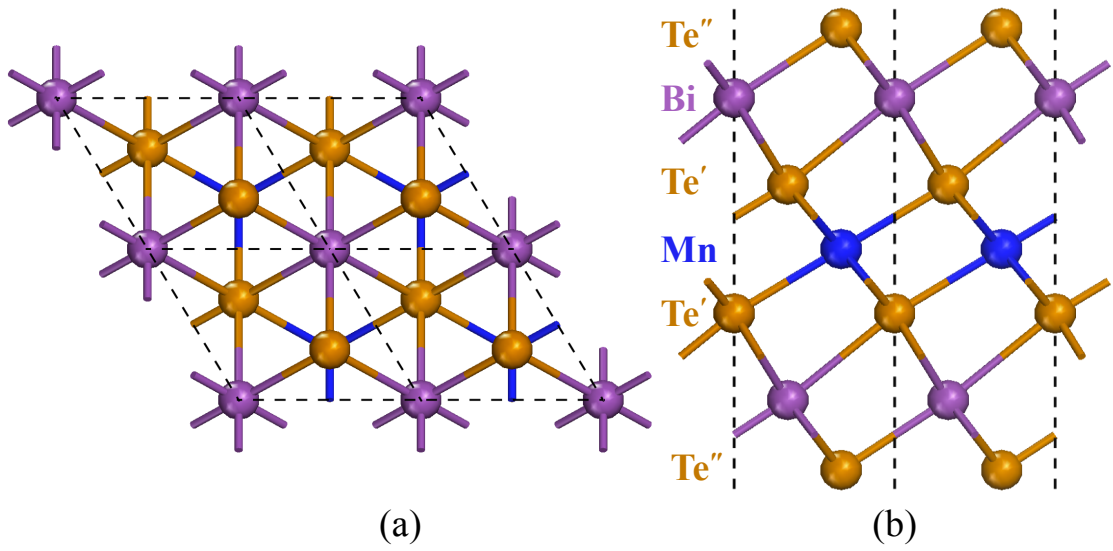

Fig. 1. (a) Top view and (b) side view of the atomic structure of monolayer $MnBi_2Te_4$. The rhombuses formed by dashed lines represent the in-plane unit cells.

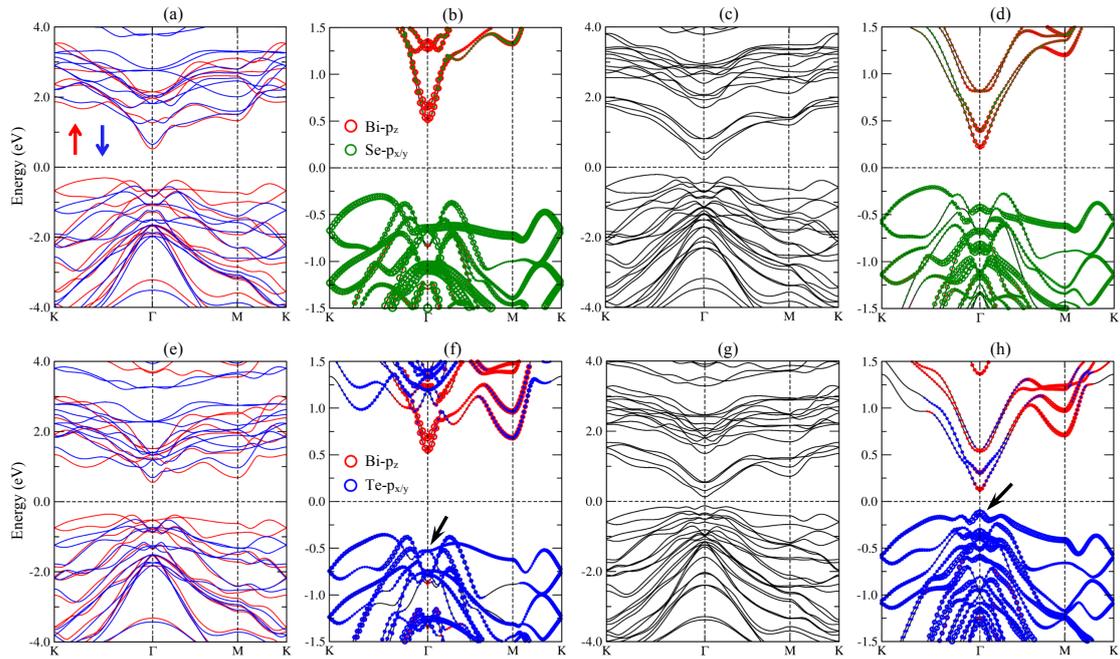

Fig. 2. Band structures of monolayer $MnBi_2Se_4$ (top panels) and $MnBi_2Te_4$ (bottom panels). (a, b, e, f) are from spin-polarized calculations without the spin-orbit coupling (SOC), while the SOC was considered in (c, d, g, h). (b, d, f, h) are orbital-resolved band structures with the weights of the Bi-$p_z$ and Se/Te-$p_{x/y}$ orbitals represented by sizes of dots. Arrows in (a) denote the majority spin (red) and minority spin (blue). The Fermi level is defined as zero energy.

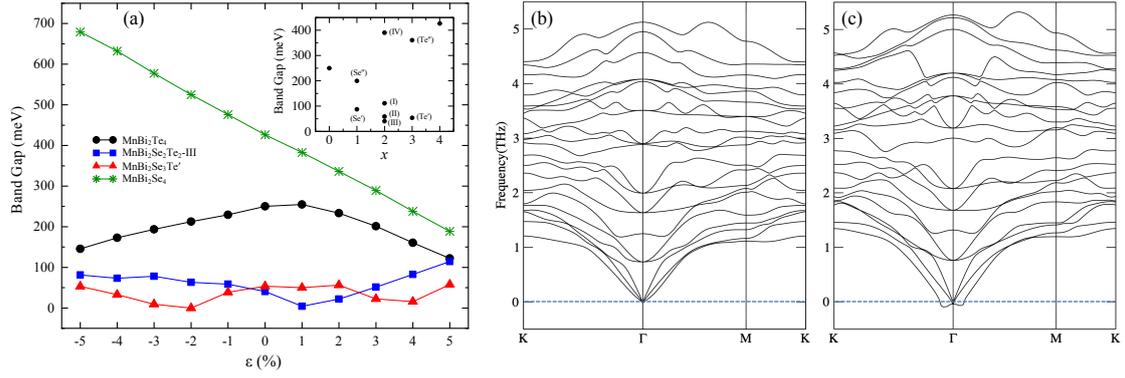

Fig. 3. (a) Band gaps (with the spin-orbit coupling) of monolayer (ML) MnBi$_2$Se$_x$Te$_{4-x}$ under biaxial strain ($\varepsilon$). The negative and positive signs of $\varepsilon$ denote compressive and extensile strains, respectively. The inset displays the band gaps as a function of x without strain. For ML MnBi$_2$SeTe$_3$ (x = 1), two configurations exist: MnBi$_2$Se'Te$_3$ and MnBi$_2$Se''Te$_3$, where the Se layer replaces an inner and outer Te layer, respectively (see Fig. 1). Similar notations are employed for ML MnBi$_2$Se$_3$Te. ML MnBi$_2$Se$_2$Te$_2$ exhibits four configurations with Se replacing: (I) the outer Te layers on both sides; (II) the outer Te layer on one side and the inner Te layer on the other side; (III) the inner and outer Te layers on the same side; (IV) the inner Te layers on both sides. (b) and (c) Phonon spectra of ML MnBi$_2$Se$_2$Te$_2$-III and MnBi$_2$Se$_3$Te'.

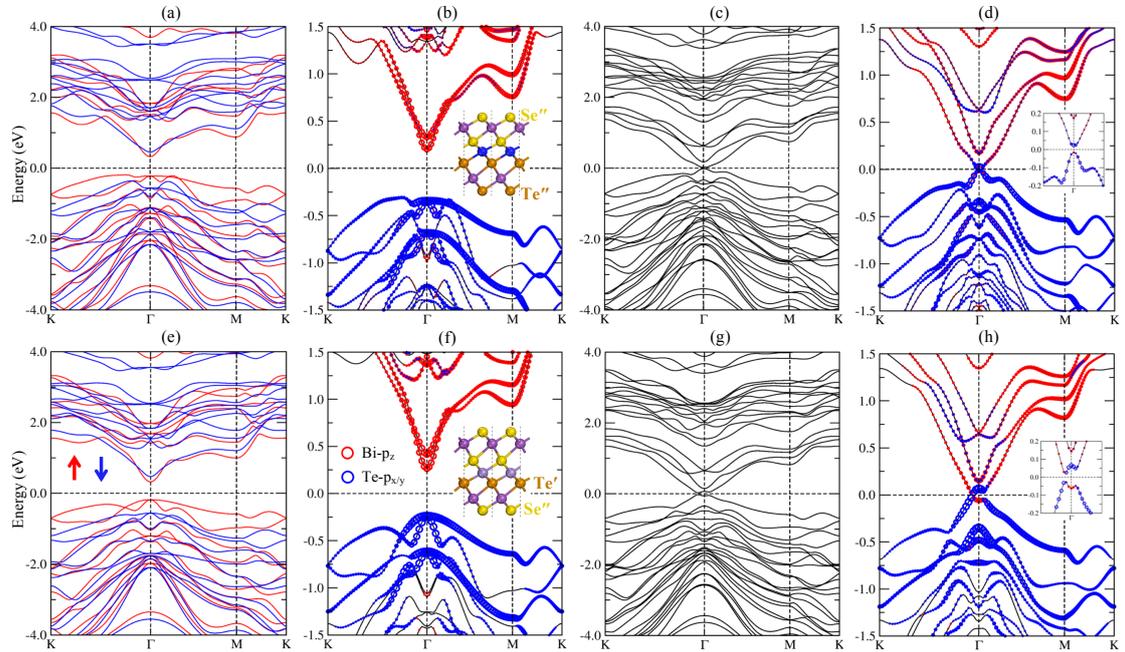

Fig. 4. Band structures of monolayer MnBi$_2$Se$_2$Te$_2$-III (top panels) and MnBi$_2$Se$_3$Te' (bottom panels). The arrangement and notations are identical to those in Fig. 2. Insets in (d) and (h) zoom in the bands near the Fermi level around the $\Gamma$ point. Insets in (b) and (f) display the side views of the atomic structures.

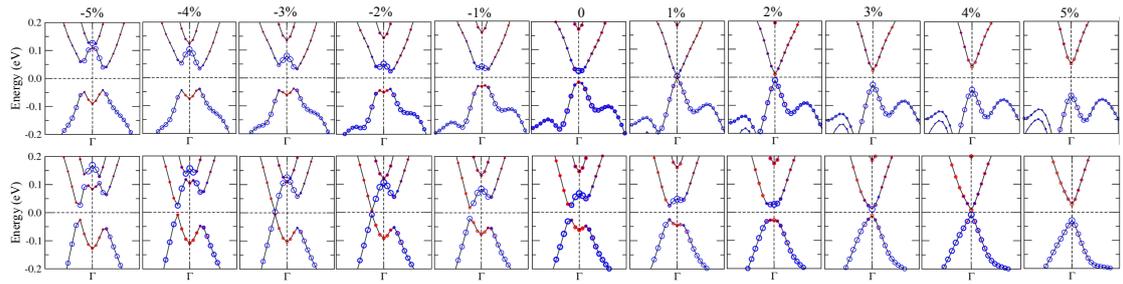

Fig. 5. Orbital-resolved band structures of monolayer MnBi$_2$Se$_2$Te$_2$-III (top panels) and MnBi$_2$Se$_3$Te' (bottom panels) under different biaxial strains, with the weights of the Bi-p$_z$ (red) and Te-p$_{x/y}$ (blue) orbitals represented by sizes of dots. The Fermi level is defined as zero energy.

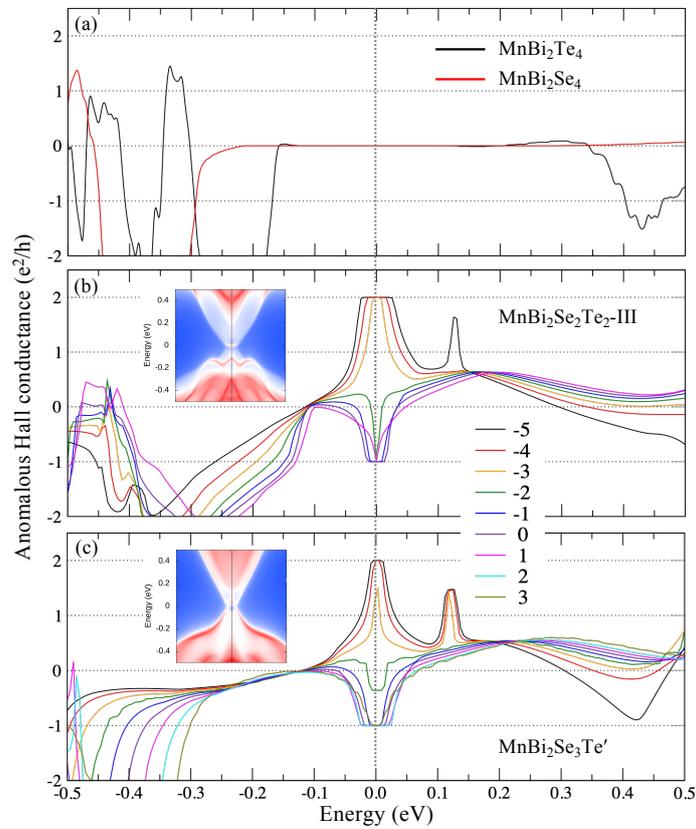

Fig. 6. Fermi level-dependent anomalous Hall conductance of four monolayer MnBi$_2$Se$_x$Te$_{4-x}$: (a) MnBi$_2$Se$_4$ and MnBi$_2$Te$_3$, (b) MnBi$_2$Se$_3$Te', (c) MnBi$_2$Se$_2$Te$_2$-III. Different colors of curves in (b,c) denote biaxial strains from -5% to 3%. The insets in (b,c) show the edge states of one-dimensional nanoribbons of monolayer MnBi$_2$Se$_3$Te' and MnBi$_2$Se$_2$Te$_2$-III. The Fermi level is defined as zero energy.

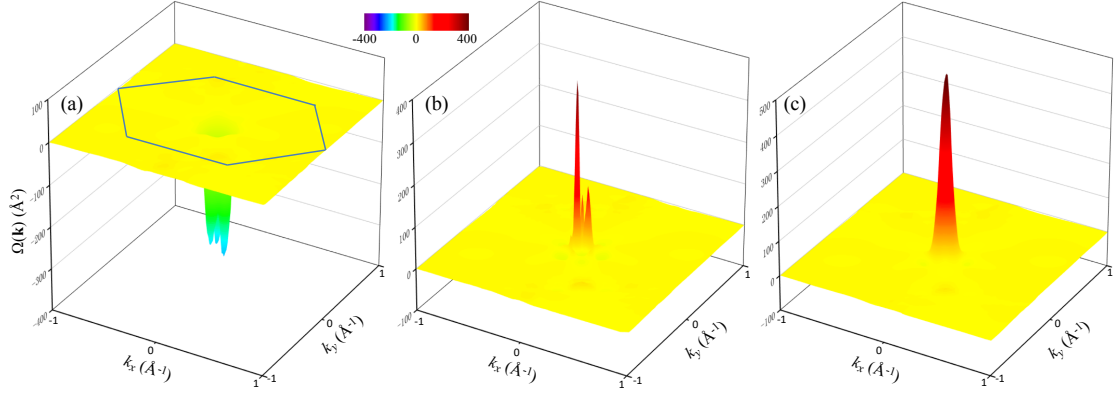

Fig. 7. The k-point dependent Berry curvature [$\Omega(\mathbf{k})$] of monolayer MnBi$_2$Se$_2$Te$_2$-III with strain (a) -5%, (b) -2%, (c) 0. The color bar represents different values of $\Omega$. The hexagon in (a) indicates the first Brillouin zone, where the six corners are the K and K' points and the center is the $\Gamma$ point.

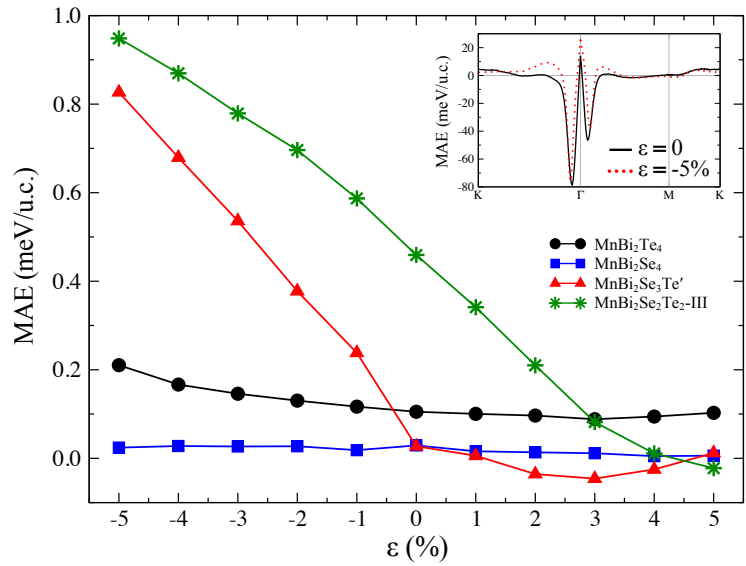

Fig. 8. Magnetocrystalline anisotropy energy (MAE) of a unit cell (u.c.) of four monolayer MnBi$_2$Se$_x$Te$_{4-x}$ under biaxial strain ($\varepsilon$). The inset shows the k point-dependent MAE of monolayer MnBi$_2$Se$_3$Te'.